# Compaction of Granular Columns under Thermal Cycling

Yuxuan Luo,[1] Haiyang Lu,[3] Xinyu Ai,[1] Zelin Liu,[1] Houfei Yuan,[3] Zhuan Ge,[1] Zhikun Zeng,[3,*]

and Yujie Wang[1,2,3,†]

[1]School of Physics, Chengdu University of Technology, Chengdu 610059, China
[2]State Key Laboratory of Geohazard Prevention and Geoenvironment Protection, Chengdu University of Technology, Chengdu 610059, China
[3]School of Physics and Astronomy, Shanghai Jiao Tong University, Shanghai 200240, China

Granular materials undergo compaction under periodic temperature fluctuations, leading to various engineering and geological phenomena from landslides to silo compaction. Although thermal effects on granular materials have been extensively studied in soil mechanics and geology, the underlying physical mechanisms remain unclear. This study investigates the compaction dynamics of granular materials subjected to thermal cycling using monodisperse glass beads and polydisperse sand packings. We demonstrate that differential thermal expansion between the container and the grains drives compaction through shear in our experimental systems. We quantify compaction dynamics using three established fitting models: Kohlrausch-Williams-Watts (KWW), double-exponential, and logarithmic functions. Our results reveal that granular materials exhibit slow relaxation processes in response to weak perturbations, displaying aging dynamics similar to those observed in glassy systems. These findings provide insights into fundamental mechanisms of granular compaction with broad implications for geological and engineering applications.

## I. INTRODUCTION

Granular materials are ubiquitous in both natural processes and industrial applications

[1,2]. In real-world scenarios, such as sand dunes or stored grain in silos, these materials often experience natural temperature fluctuations that can induce subtle variations in grain sizes, generating significant changes in mechanical stability and flow properties [3,4]. These thermally induced deformations directly correspond to various geological processes, including thermal collapse in soils [5], and the initiation of landslides or avalanches [6], as well as engineering challenges like unwanted silo compaction [7]. Given this practical significance, researchers in soil mechanics and geology have extensively studied the thermal effects on granular materials across different soil compositions [8], particle size distributions [9,10], and boundary conditions [11,12], with particular emphasis on irreversible compaction processes under thermal cycling [5,8-12]. However, understanding granular materials' diverse and sometimes contradictory volume change behaviors under temperature variations remains a major challenge, primarily due to a limited fundamental understanding of the underlying physical mechanisms [5].

Granular materials under cyclically varying temperatures experience a systematic increase in the overall density [13]. Recent studies have demonstrated that this thermally-induced compaction process is closely related to internal stress fluctuations and creep motion with granular systems during temperature changes [14-16]. A key open question is whether these thermal effects are primarily driven by extrinsic factors, such as the thermal dilation of the container, or intrinsic ones like the thermal response of the grains. A systematic experimental investigation into these mechanisms is essential for developing a comprehensive understanding of how temperature variations affect granular pack evolution. Additionally, this thermal cycling-induced compaction of granular materials shares notable similarities with the compaction and slow relaxation of granular systems subjected to consecutive mechanical energy inputs, such as tapping and cyclic shear [17-22]. Over recent decades, extensive experimental and theoretical studies have investigated the compaction dynamics of granular systems [19-21]. When energy is applied, these systems evolve toward a steady state with a characteristic volume fraction that depends on perturbation intensity but remains independent of the initial packing configurations [21,22]. This compaction process exhibits slow dynamics akin to those observed in glassy systems. In particular, under sufficiently weak perturbations, relaxation becomes so gradual that these systems cannot reach their steady state within experimental timescales, and the response of a granular pile resembles aging phenomena in glasses [23]. This raises fundamental questions regarding whether thermally driven granular compaction displays aging characteristics similar to those observed under tapping or cyclic shear conditions.

In this paper, we investigate the compaction process of an elongated column of granular packings subjected to periodic temperature variations. By comparing the evolution of granular columns' top heights and the simultaneous deformation of the container during a single thermal cycle, we demonstrate that the difference in the thermal expansion between the container and the grains primarily induces compaction in our system. This disparity enables internal grains to experience cyclic shear from the container walls with an extremely small amplitude. Subsequently, we fit the compaction curves of volume fraction for both monodisperse spherical particles and polydisperse sand using various fitting formulas. Our results show that the thermal cycling-induced compaction process in granular columns follows aging dynamics commonly observed in glassy systems. These findings advance our fundamental physical understanding of thermal effects in granular systems and offer new insights into the development of relevant theories in engineering and soil mechanics, which may provide valuable guidance for their practical applications.

## II. EXPERIMENTAL METHODS

To investigate the compaction process of granular materials under thermal cycling, we periodically heat and cool a column of granular packings within an elongated cylindrical container [Fig. 1(a)]. We utilize two types of grains: monodisperse sodium-calcium glass beads with a diameter of $d = 3$ mm and a thermal expansion coefficient of $8.6 \times 10^{-6}$ K$^{-1}$, as well as standard sand with a particle size range of 0.08 to 2 mm and its associated thermal expansion coefficient of $5.5 \times 10^{-7}$ K$^{-1}$. The cylindrical container is made of PVC plastic that can withstand high temperatures and is fixed vertically to the ground by aluminum profiles. It is noteworthy that the thermal expansion coefficient of the PVC materials, $\lambda = 2.7 \sim 4.4 \times 10^{-5}$ K$^{-1}$, is at least 3 times greater than those of the grains. The container has a height of $l = 1.2$ m, an inner diameter of $D = 22$ mm, and a wall thickness of $w = 2$ mm. The elongated shape of the pipe is designed to enhance the visibility of height changes in granular packings during the thermal cycling process. A circular marker is affixed to the pipe at a height of 1.08 m to monitor the deformation of the container throughout thermal cycling. Before conducting experiments, granular grains are poured into the container, which is then turned upside down repeatedly to obtain an initially loose packing. The initial height of this packing measures approximately 1.14 m, slightly exceeding the marker's position to avoid visual occlusion.

The temperature of the granular pile is measured using a temperature sensor (Platinum Pt100) placed at the bottom of the pile. Heating is accomplished with a heating cable (Voltage: 220V; Power: 4500 W/m²; Temperature range: -200 to 400°C) wrapped around the PVC pipe extending from the base to a height of 1 m, providing nearly uniform heat distribution along the pipe. Consequently, deformation of the container occurs within the lower section of the pipe with height $l_{heat} = 1\,\text{m}$. The heating and cooling processes alternate based on the system temperature reaching predetermined thresholds. Each thermal cycle involves heating to preset temperatures $T_{max}$, followed by natural cooling to $T_{min} = 30°C$, which is slightly above room temperature (23°C). A full thermal cycle lasts about 40 minutes, with heating taking about 16 minutes and cooling requiring about 24 minutes. For monodisperse glass beads, we applied temperature variations of $\Delta T = T_{max} - T_{min} = 20°C$, 30°C, 40°C, and 50°C, while for standard sand, $\Delta T = 30°C$. We perform about 200 cycles for each temperature condition to study the resulting granular compaction, with denoting cycle number as $n$.

Simultaneously, the top height of the granular pile and the circular marker are recorded using a camera (Hikvision, MV-CU103-A0GM) that captures images every 5 seconds. A light source is positioned opposite to the camera to enhance the contrast ratio of the images. Figure 1(b) shows a sample image with its bottom at a height of $h_b = 1.065\,\text{m}$ and an area of 10.1 cm (height) × 4.8 cm (width). The height of the granular column is determined through subsequent image processing techniques. After binarizing the initial image, we apply standard erosion and top-hat transformation methods to eliminate pipe wall pixels [24]. By dividing the total area occupied by granular packing pixels by their corresponding widths, we can accurately obtain the top height of the granular column from the bottom of the image with an error margin of less than 0.15 mm. Additionally, we simultaneously track the height evolution of the circular markers. By adding these measurements to the images' bottom height, we can obtain the total height of the granular packing $h$ and the marker $h_m$, respectively.

## III. RESULTS

### A. Compaction mechanism under thermal cycling

Figure 2 shows the height evolution of both the monodisperse glass bead and polydisperse sand packings under thermal cycling. While the heights of the granular column experience a periodical variation during the thermal cycling, they show a continuous overall decline, indicating compaction induced by thermal cycling. Notably, the degree of compaction is more pronounced in glass bead systems compared to sand systems, with the compaction rate increasing as the temperature difference $\Delta T$ grows. Specifically, Figures 3(a) and 3(b) show the height evolution of both the granular pile composed of monodisperse glass beads and the marker during the initial 20 thermal cycles with a temperature variation of $\Delta T = 30°C$. Throughout each thermal cycle, the height of the marker exhibits an increase during heating followed by a decrease during cooling, and it returns to its original height after completing a full cycle. Specifically, the marker height variations within each cycle follow an exponential form during both the heating and cooling phases [inset of Fig. 3(b)]. Given that pipe length varies linearly with temperature, this behavior can be attributed to constant heat input from the heating cable alongside a dissipation process whose rate is proportional to the temperature difference from room temperature. The thermal expansion coefficient of the PVC pipe can be estimated by $\lambda = \frac{\Delta h_m / l_{heat}}{\Delta T} \approx \frac{1.3 \text{ mm} / 1000 \text{ mm}}{30 \text{ K}} = 4.3 \times 10^{-5} \text{ K}^{-1}$, where $\Delta h_m$ represents the maximum change in marker height over a period and $l_{heat}$ is the length of heated pipe undergoing deformation. The experimentally measured value of $\alpha$ roughly matches the standard measurement results, validating our method for assessing pipe deformation based on variations in marker height.

A notable observation is that despite the grains' thermal expansion coefficient being much smaller than that of the PVC pipe, the height change of the granular column (~1.5 mm) is comparable to the pipe deformation. This significantly exceeds the estimated height variation of ~0.3 mm [black dotted curve in the inset of Fig. 3(a)] that results from particle deformation alone within a single cycle. This suggests that container wall deformation primarily drives the height changes in the granular column. In summary, the differential thermal expansion between the container and grains induces cyclic shear on the granular packings from the container wall, with a corresponding shear amplitude of $\Gamma = \frac{\Delta h_m/2}{h_{heat}} \approx \frac{1.3 \text{ mm}/2}{1000 \text{ mm}} = 6.5 \times 10^{-4}$. This shear differs from simple shear geometry due to the linear variation of pipe deformation along its height. Since the deformation of the container wall is proportional to temperature changes, different temperature variations can be interpreted as corresponding to different shear amplitudes.

### B. Relaxation dynamics of granular packings

From the height evolution of the granular column, we can calculate the corresponding evolution of the packing's volume fraction. We measure the total volume of the grains using drainage methods. Randomly selected particles with a mass of 100 g are introduced into a measuring cylinder containing water, with the water level rising corresponding to the volume occupied by these grains. This measurement is repeated 10 times for statistical reliability. The total grain mass is then measured and converted to total volume $V_{grain}$. The volume fraction of the granular column is calculated by $\Phi = V_{grain} / \left[ \pi \left( \frac{D}{2} \right)^2 h \right]$. As shown in Figs. 4(a-d), the granular columns compact under thermal cycling.

In order to quantify the compaction process, we use three fitting models that are commonly employed to analyze either compaction in granular materials or relaxation dynamics in glassy systems. The first model is the Kohlrausch-Wiliams-Watts (KWW) function:

$$\phi(\delta\gamma) = \phi_f - (\phi_f - \phi_0) \exp\left(-\frac{n}{\tau}\right)^\beta, \tag{1}$$

where $\phi_f$ is the steady-state packing density, $\phi_0$ is the initial volume fraction, $\tau$ denotes the relaxation time, and $\beta$ is a fitting parameter that captures the heterogeneous relaxation across the material. This function describes the slow relaxation process as a system transitions from out-of-equilibrium toward equilibrium. The fitting parameter $\beta$ is typically less than 1, indicating that the relaxation occurs in a stretched or delayed manner compared to simple exponential decay. This stretched exponential form captures a distribution of relaxation times rather than a singular timescale within disordered and heterogeneous systems [25,26].

The second approach is the double-exponential (D-exp) fitting:

$$\phi(\delta\gamma) = \phi_0 - A_1 \exp\left(-\frac{n}{\tau_1}\right) - A_2 \exp\left(\frac{n}{\tau_2}\right), \tag{2}$$

where $\phi_0$ is the initial volume fraction, $\tau_1$ and $\tau_2$ are two relaxation timescales, and $A_1$ and $A_2$ are fitting parameters. This model suggests that the system's relaxation behavior is governed by two independent processes occurring at different timescales [27-30]. In granular systems, these processes may correspond to the local relaxation of individual particles at shorter timescales and multi-particle relaxation at longer timescales.

The final model is the logarithmic (log) fitting:

$$\phi(\delta\gamma) = \phi_f - \frac{\Delta\phi_\infty}{1 + B\ln\left[1 + \frac{n}{\tau}\right]}, \tag{3}$$

where $\phi_f$ is the maximum packing density achievable by the system, $\phi_\infty$ is the density variation during the whole relaxation process, $\tau$ is the relaxation time, and $B$ is a fitting parameter. The logarithmic form is frequently used to characterize slow relaxation dynamics in out-of-equilibrium states, such as during aging processes [21,31,32]. This model decays significantly more gradually than exponential models, particularly as the system is very close to its steady state.

Figures 4(a-d) present the fitting results of three models for monodisperse glass bead systems at different $\Delta T$. The calculated relaxation times for different $\Delta T$ are shown in Fig. 5(a), demonstrating a decrease as temperature variation increases, as expected. The relaxation time obtained from the logarithmic fitting is much larger than those derived from the other two methods, which is attributed to the reason for the specific fitting form. Additionally, in the double-exponential fit, two distinct relaxation timescales are identified, indicating the complex nature of relaxation within granular systems. Subsequently, we quantify fitting accuracy using mean square error (MSE) $\frac{1}{N-1}\sum_n\left[\Phi(n) - f(n)\right]^2$, where $N$ is the number of experimental data points, and $\Phi(n)$ and $f(n_i)$ denote the measured and predicted packing fractions at the $n$th cycle, respectively. As shown in Fig. 5(b), the MSE for the KWW fit is larger compared to those obtained from both double-exponential and logarithmic fits, which exhibit comparable values. Considering that the double-exponential form introduces too many fitting parameters, we believe that the logarithmic fit, which represents the typical aging dynamics in glasses, is the most appropriate. This is consistent with the observation in tapped granular systems, wherein relaxation dynamics favor KWW fits at high intensities when systems can reach their steady state, but follow logarithmic fits under weak tap intensities. This suggests that the compaction dynamics of granular systems under thermal cycling may share the same fundamental physical mechanism as those observed under weak tapping. We hypothesize that this phenomenon arises from broken ergodicity in systems subjected to weak perturbations, whereby these systems remain confined to local energy minima in the energy or free-energy landscape, analogous to aging processes [33,34].

Furthermore, the compaction process in standard sand systems can, in principle, be described by these three models, though it exhibits discontinuities rather than the smooth evolution observed in glass bead systems, corresponding to the sudden avalanche behavior [Fig. 6(a)]. We also note that the compaction process of sand systems is slower than that of glass beads [solid markers in Fig. 6(b)]. We attribute these differences to the inherent polydispersity in sand particles, where compaction occurs through smaller particles navigating through gaps between larger particles, rather than through straightforward configurational rearrangements. These more intricate relaxation mechanisms slow down the compaction process in sand systems.

## IV. CONCLUSION

In conclusion, this study provides a comprehensive investigation into the compaction dynamics of granular materials subjected to thermal cycling. We have demonstrated that, in our experimental systems, the primary mechanism driving thermal cycling-induced compaction is the differential thermal expansion between the container and the granular grains, which induces cyclic shear on the internal grains. By employing various fitting models, we have quantitatively characterized the compaction behavior of both monodisperse glass beads and polydisperse sands. Our results reveal that the compaction process of granular materials exhibits characteristics similar to those observed in aging dynamics within glassy systems, underscoring the extremely slow relaxation of granular materials under weak perturbations. The findings shed new light on the role of thermal effects in granular systems and offer valuable insights into the fundamental mechanisms of granular compaction and relaxation. Such insights could advance existing research methods in soil mechanics and engineering primarily based on empirical phenomena, and hold significant implications for understanding related real-world processes.

## ACKNOWLEDGMENTS

The work is supported by the National Natural Science Foundation of China (No. 12274292) and the Space Application System of China Manned Space Program (KJZ-YY-NLT0504).

Corresponding author

*zzk97115_kenny@sjtu.edu.cn

†yujiewang@sjtu.edu.cn

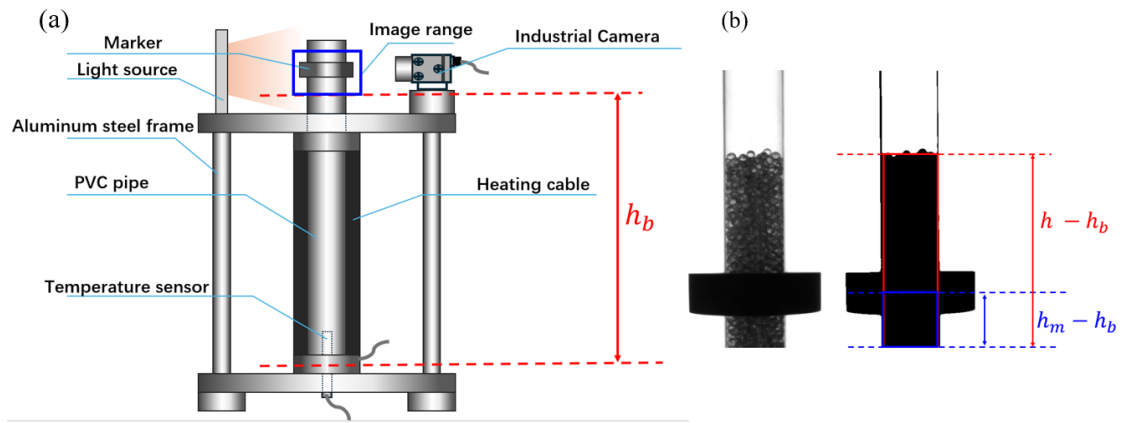

FIG. 1. (a) Schematic diagram of the experimental setup, with the camera capturing the area within the blue frame. (b) Raw image (left) and binarized image (right) of a monodisperse glass bead column under thermal cycling.

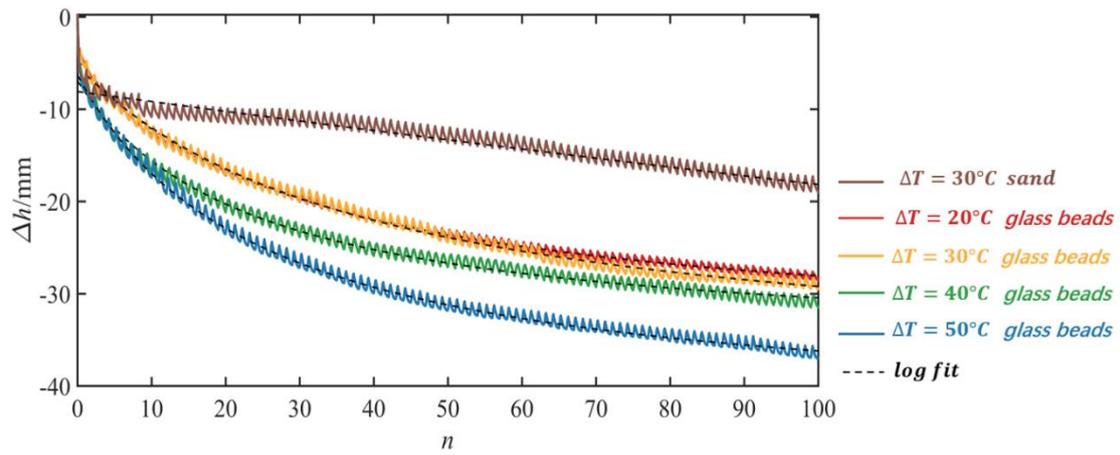

FIG. 2. (a) Height variations $\Delta h$ of glass bead and sand packings as a function of the number of thermal cycles *n* under different temperature variations $\Delta T$. The black dotted curves denote the logarithmic fits.

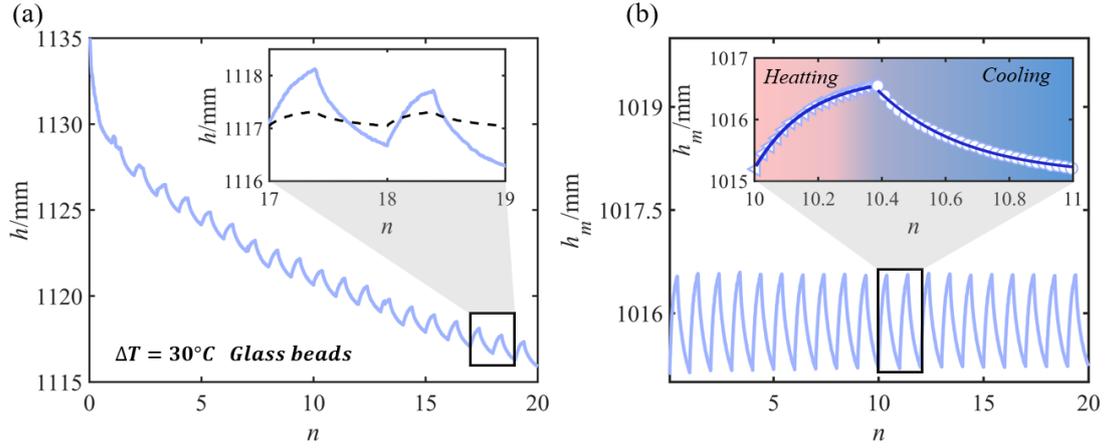

FIG. 3. (a) Height evolution of a glass bead packing $h$ as a function of the number of thermal cycles $n$ under a temperature variation of $\Delta T = 30°C$. Inset: magnified portion of the compression data showing the compaction curve. The black dotted curve corresponds to the estimated height variation of glass bead packings due to the thermal dilatancy alone. (b) Height evolution of markers $h_m$ as a function of the number of thermal cycles $n$. Inset: magnified portion of the compression data showing the compaction curve. Blue solid curves denote the single exponential fitting for the heating and cooling stages, $A + B\exp\left(-\dfrac{t}{\tau}\right)^{\beta}$, respectively.

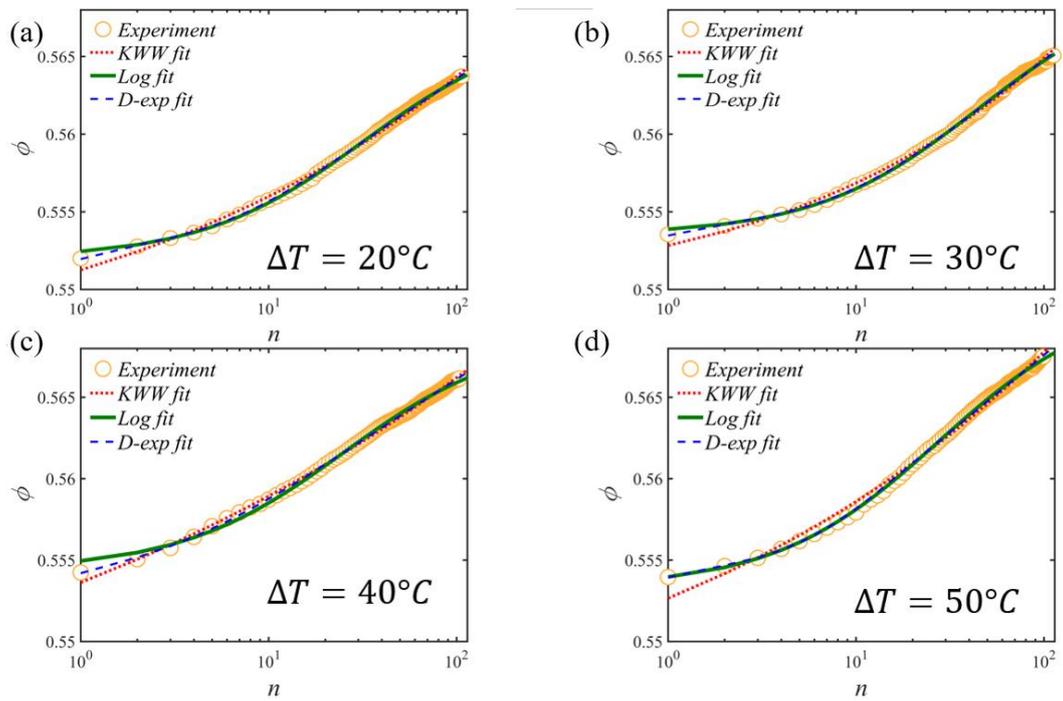

FIG. 4. Volume fractions $\phi$ as a function of the number of thermal cycles $n$ at $\Delta T =$ (a) 20°C, (b) 30°C, (c) 40°C and (d) 50°C. The dotted red curves, green solid curves, and dotted blue curves are the KWW, logarithmic, and double-exponential fittings, respectively.

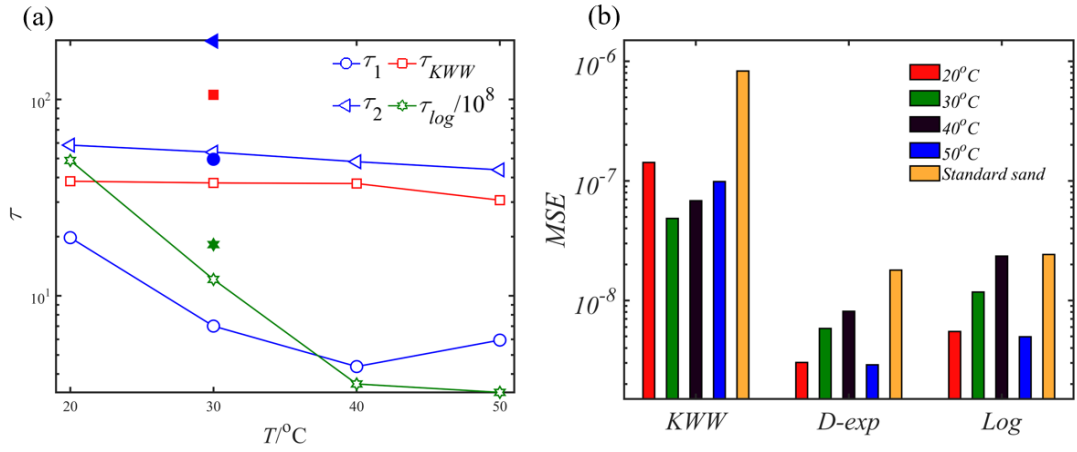

FIG. 5. (a) Relaxation timescale obtained by different fitting methods for both glass bead packings (hollow markers) and sand packings (solid markers) at different temperature variations. (b) MSE obtained from the three fitting methods.

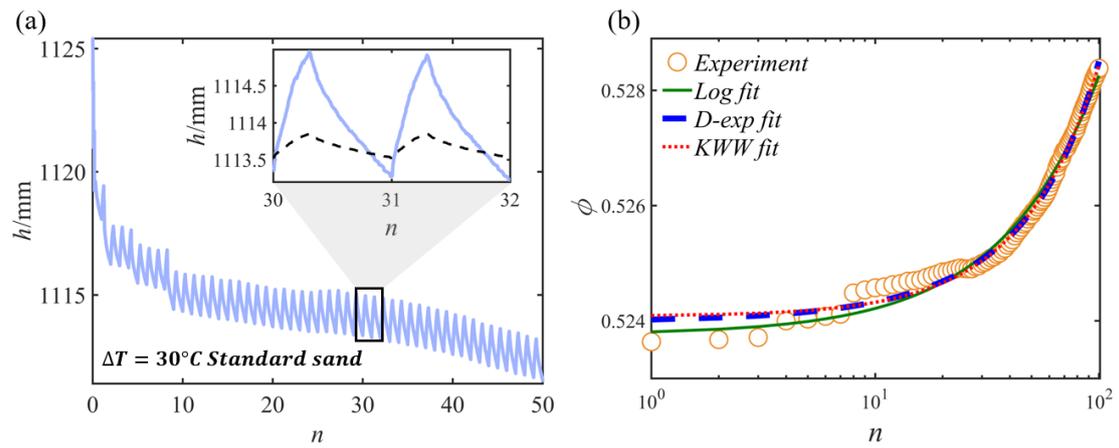

FIG. 6. (a). Height evolution $h$ of a sand packing as a function of the number of thermal cycles $n$ under a temperature variation of $\Delta T = 30°C$. Inset: magnified portion of the compression data showing the compaction curve. (b) Volume fractions as a function of the number of thermal cycles.